\documentclass{article}
\usepackage{amssymb}
\usepackage{amsfonts}
\usepackage{graphicx}
\usepackage{amsmath}
\usepackage{geometry}
\usepackage{scalefnt}
\usepackage{ulem}
\usepackage{hyperref}

\input{tcilatex}

\begin{document}

\title{Observables of Angular Momentum as Observables on the Fedosov
Quantized Sphere}
\author{Philip Tillman$^{1}$, George Sparling$^{2}$ \\
$^{1}$Department of Physics and Astronomy, University of Pittsburgh,
Pittsburgh, PA, USA\\
$^{2}$Department of Mathematics, University of Pittsburgh, Pittsburgh, PA,
USA\\
email:$^{1}$phil.tillman@gmail.com $^{2}$sparling@twistor.org}
\date{\today }
\maketitle

\begin{abstract}
In this paper we construct quantum mechanical observables of a single free
particle that lives on the surface of the two-sphere $\mathbb{S}^{2}$ by
implementing the Fedosov $\ast $-formalism. The Fedosov $\ast $ is a
generalization of the Moyal star product on an arbitrary symplectic
manifold. After their construction we show that they obey the standard
angular momentum commutation relations in ordinary nonrelativistic quantum
mechanics. The purpose of this paper is three-fold. One is to find an exact,
non-perturbative solution of these observables. The other is to verify that
the commutation relations of these observables correspond to angular
momentum commutation relations. The last is to show a more general
computation of the observables in Fedosov $\ast $-formalism; essentially an
undeformation of Fedosov's algorithm.
\end{abstract}

\section{Introduction}

The Moyal star product formalism is an equivalent way to do quantum
mechanics.$\left[ \text{\hyperlink{3}{3}}\right] $ \ The idea is that
instead of using abstract linear operators on a Hilbert space such as
position $\hat{x}$ and momentum $\hat{p}$, we may use classical variables $x$
and $p$ however we change the product so that the commutation relations are
the same as in the Hilbert space formalism. \ Namely:%
\begin{equation*}
\left[ \hat{x}^{a},\hat{p}_{b}\right] =i\hbar \delta _{b}^{a}\text{ \ \ },%
\text{ \ \ }\left[ \hat{x}^{a},\hat{x}^{b}\right] =0=\left[ \hat{p}_{a},\hat{%
p}_{b}\right]
\end{equation*}%
become:%
\begin{equation*}
\left[ x^{a},p_{b}\right] _{\ast }=i\hbar \delta _{b}^{a}\text{ \ \ },\text{
\ \ }\left[ x^{a},x^{b}\right] _{\ast }=0=\left[ p_{a},p_{b}\right] _{\ast }
\end{equation*}%
we use the convention that the lower case indices run from $1,\ldots ,n$ and
capital ones run from $1,\ldots 2n$ and: 
\begin{equation*}
\left[ f,g\right] _{\ast }=f\ast g-g\ast f
\end{equation*}%
where $f$ and $g$ are any 2 functions of $x$ and $p$.

We note that the limit $\hbar \rightarrow 0^{+}$ gives the ordinary product
of functions.

The definition of the Moyal star for $\mathbb{R}^{2n}$\ explicitly is:%
\begin{equation*}
f\ast g=fe^{\frac{i\hbar }{2}\omega ^{AB}\overleftarrow{\partial }_{A}%
\overrightarrow{\partial }_{B}}g=fg+\frac{i\hbar }{2}\omega ^{AB}\left(
\partial _{A}f\right) \left( \partial _{B}g\right) -\frac{\hbar ^{2}}{8}%
\omega ^{CE}\omega ^{AB}\left( \partial _{C}\partial _{A}f\right) \left(
\partial _{E}\partial _{B}g\right) +\cdots
\end{equation*}%
where $\partial _{A}=\left( \frac{\partial }{\partial x^{a}},\frac{\partial 
}{\partial p_{a}}\right) $ and the arrow determines the direction that the
derivative acts and the operator $\omega ^{AB}\overleftarrow{\partial }_{A}%
\overrightarrow{\partial }_{B}$\ is called the Poisson bracket.

There is an invertible map called the Weyl transform $\mathcal{W}$ that
translates from the Hilbert space formalism to the Moyal formalism.\ The
main property of this transform is that an arbitrary Taylor series operator
on the Hilbert space:\footnote{%
Note that this is effectively an arbitrary operator since we can use the
commutators to rearrange each term so that the $x$'s are to the left and the 
$p$'s are to the right.}%
\begin{equation*}
\hat{A}=\sum_{m,n}A_{a_{1}\cdots a_{m}}^{~~~~~~~~b_{1}\cdots b_{n}}\hat{x}%
^{a_{1}}\cdots \hat{x}^{a_{m}}\hat{p}_{b_{1}}\cdots \hat{p}_{b_{n}}
\end{equation*}%
becomes by applying the Weyl transform:%
\begin{equation*}
\mathcal{W}\left( \hat{A}\right) =A=\sum_{m,n}A_{a_{1}\cdots
a_{m}}^{~~~~~~~~b_{1}\cdots b_{n}}x^{a_{1}}\ast \cdots \ast x^{a_{m}}\ast
p_{b_{1}}\ast \cdots \ast p_{b_{n}}
\end{equation*}%
in a mechanical way by simply replacing each $\hat{x}$ with $x$, $\hat{p}$
with $p$ and placing stars between each of them as is done above.$\left[ 
\text{\hyperlink{3}{3}}\right] $

The trace over an operator of compact support goes to:%
\begin{equation*}
Tr\left( \hat{A}\right) \overset{\mathcal{W}}{\leftrightarrow }Tr_{\ast
}\left( A\right) :=\frac{1}{\left( 2\pi \hbar \right) ^{n}}\int \frac{\omega
^{n}}{n!}A
\end{equation*}%
So if we are given the Hamiltonian $\hat{H}$ and the density matrix $\hat{%
\rho}$ we may map:%
\begin{equation*}
\hat{H}\overset{\mathcal{W}}{\leftrightarrow }H\text{ \ , \ \ }\hat{\rho}%
\overset{\mathcal{W}}{\leftrightarrow }\rho
\end{equation*}%
We thus can get the time-independent Schr\"{o}dinger equation by mapping:%
\begin{equation*}
\hat{H}\hat{\rho}_{n}=E_{n}\hat{\rho}_{n}\text{ \ \ , \ \ }\left[ \hat{H},%
\hat{\rho}_{n}\right] =0
\end{equation*}%
to:%
\begin{equation*}
H\ast \rho _{n}=E_{n}\rho _{n}\text{ \ \ , \ \ }\left[ H,\rho _{n}\right]
_{\ast }=0
\end{equation*}%
where $\rho _{n}$\ are called the Wigner functions.\ This also works with
the time-dependent Schr\"{o}dinger equation.\footnote{%
See Fedosov for clarification.$\left[ 1\right] $}

Also expectation values become:%
\begin{equation*}
Tr\left( \hat{\rho}\hat{A}\right) \leftrightarrow Tr_{\ast }\left( \rho \ast
A\right)
\end{equation*}

The Moyal $\ast $ has been generalized to an arbitrary smooth symplectic
manifold $\left( \mathcal{N},\omega ,D\right) $ endowed with a preserved
two-form $\omega $ (called the symplectic form) and a phase-space connection 
$D$ by Fedosov.$\left[ \text{\hyperlink{1}{1}}\right] $(an excellent summary
is $\left[ \text{\hyperlink{2}{2}}\right] $) For any such manifold $\left( 
\mathcal{N},\omega ,D\right) $ he gives a perturbative expansion for his $%
\ast $-product. However, the convergence issues of the Fedosov $\ast $, in
general, remain unknown.

The properties of the Fedosov $\ast $ are:

\begin{itemize}
\item It is an associative (but not commutative) map $\ast :C^{\infty
}\left( \mathcal{N}\right) \times C^{\infty }\left( \mathcal{N}\right)
\rightarrow C^{\infty }\left( \mathcal{N}\right) $.

\item Invariant under \underline{all} smooth coordinate transformations of
the phase-space variables $x$ and $p$.

\item No assumed Hamiltonian.

\item The Fedosov $\ast $ is given perturbatively given any symplectic
manifold $\left( \mathcal{N},\omega ,D\right) $.

\item In the limit $\hbar \rightarrow 0^{+},$ $\ast $ becomes the ordinary
pointwise multiplication of functions on $\mathcal{N}$.

\item To first order in $\hbar $ the commutator is the Poisson bracket: $%
\left[ f,g\right] _{\ast }=i\hbar \left\{ f,g\right\} +\mathcal{O}\left(
\hbar ^{2}\right) $.

\item When $\mathcal{N}=T^{\ast }\mathbb{E}^{n}$ (i.e. the phase space or\
the cotangent bundle of $\mathbb{E}^{n}$)\footnote{%
Here $\mathbb{E}^{n}$ stands for Euclidean $n$-dimensional space.} we get
the Moyal $\ast $.
\end{itemize}

In this paper we restrict $\mathcal{N}$ to be the cotangent bundle of a
manifold with metric $g$ $\left( \mathcal{M},g\right) $ denoted $T^{\ast }%
\mathcal{M}$.\footnote{%
The cotangent bundle of any manifold is known to be a symplectic manifold.}\
The reason to do this is that the cotangent bundle of a manifold is the
phase-space of that manifold (i.e. the space of all coordinates $x$ and
momentum $p$). In quantum mechanics using the Moyal $\ast $ the phase-space
is the arena for quantization by giving proper $\ast $-commutation relations
between the $x$'s and $p$'s. The importance of the Fedosov $\ast $-formalism
is that it is a coordinate invariant way of constructing these commutation
relations on general $T^{\ast }\mathcal{M}$ in such a way that they patch
consistently to any coordinate map of the cotangent bundle. Also another
important point is that it can be constructed at least perturbatively for
any cotangent bundle.

However unlike Fedosov who defines a formulation based on the deformation of
covectors (i.e. covectors equipped with a Moyal-like product between them)
we will not. We will introduce a Heisenberg algebra generated by $\tilde{s}$
and $\tilde{k}$ ($\left[ \tilde{s}^{i},\tilde{s}^{j}\right] =\left[ \tilde{k}%
_{i},\tilde{k}_{j}\right] =0,~\left[ \tilde{s}^{i},\tilde{k}_{j}\right]
=i\hbar \delta _{j}^{i}$ where $i$ and $j$ run from $1$ through $2n$) at
every point of our phase-space $T^{\ast }\mathcal{M}$. The motivation to do
this instead of Fedosov's way is to make a more direct connection between
ordinary quantum mechanics involving Heisenberg algebras and the state
spaces that the algebra acts on called Hilbert spaces. We then define this
algebra to be linear operators on a Hilbert space which, of course, will
eventually contain our states. This new construction will still preserve all
of the essential properties of the original Fedosov $\ast $ albeit
reformulated so as to apply to different objects. It will \ be a
quantization procedure i.e. a map of the variables on the phase-space $x$
and $p$ to the observables $\hat{x}$ and $\hat{p}$ which are linear
operators on the Hilbert space.

The properties of the Fedosov $\ast $-quantization in our construction are:

\begin{itemize}
\item $\hat{x}$ and $\hat{p}$ form an associative but noncommutative algebra.

\item The map from $\left( x,p\right) \rightarrow \left( \hat{x},\hat{p}%
\right) $ is invariant under \underline{all} smooth canonical coordinate
transformations of the phase-space variables $x$ and $p$.

\item No assumed Hamiltonian.

\item We can construct the $\hat{x}$ and $\hat{p}$ perturbatively given any $%
\left( T^{\ast }\mathcal{M},\omega ,D\right) $.

\item In the limit $\hbar \rightarrow 0^{+},$ $\hat{x}$ and $\hat{p}$ become 
$x$ and $p$ respectively i.e. the ordinary variables on $T^{\ast }\mathcal{M}
$.

\item To first order in $\hbar $ the commutator is the Poisson bracket: $%
\left[ \hat{f},\hat{g}\right] =i\hbar \left\{ \hat{f},\hat{g}\right\} +%
\mathcal{O}\left( \hbar ^{2}\right) $.

\item When $\mathcal{M}=\mathbb{%
\mathbb{R}
}^{n}$ we get the ordinary quantum mechanics.
\end{itemize}

In the present work we take as our symplectic manifold$\mathcal{\ }T^{\ast }%
\mathbb{S}^{2}$, the phase space of a single particle on the 2-sphere, $%
\mathbb{S}^{2}$. For this space we construct the Fedosov observables
non-perturbatively. The advantage of choosing $\mathbb{S}^{2}$ is that we
had suspected previous to the calculation that the commutators are the same
as the usual angular momentum commutators in nonrelativistic quantum
mechanics. Saying in fact that the theory of angular momentum is the
quantization of the two-sphere without the need for it to be embedded in $%
\mathbb{%
\mathbb{R}
}^{3}$.

\subsection{Outline}

We will follow the basic scheme of keeping derivations sufficiently general
so as to apply to a completely general manifold with metric $\left( \mathcal{%
M},g\right) $ and then state results from our specific case of the sphere.

In section 2 we introduce the phase-space connection. We introduce the basis
of covectors of matrices/operators $\hat{y}^{A}$ on the cotangent bundle in
section 3. In section 4 we attempt to motivate and solve for a new
derivation $\hat{D}$. Also we talk a bit about $\hat{D}$'s\ ambiguities.
Moving into section 5 we explicitly compute the quantities $\hat{x}$ and $%
\hat{p}$. In section 6 we compute the commutators $\left[ \hat{x}^{a},\hat{x}%
^{b}\right] ,~\left[ \hat{x}^{a},\hat{p}_{b}\right] $ and $\left[ \hat{p}%
_{a},\hat{p}_{b}\right] $ using the explicit forms of the operators. Section
7 explains how one would construct states of angular momentum on $T^{\ast }%
\mathbb{S}^{2}$ by finally introducing the standard Hamiltonian in ordinary
nonrelativistic quantum mechanics. Up until this point no Hamiltonian was
assumed.

\section{The Phase-Space Connection for $T^{\ast }\mathbb{S}^{2}$}

Before we begin, we note the use of the convention that the lower case are
the indices of $\mathcal{M}$ (these run from $1,\ldots ,n$) and capital ones
are the indices of the phase-space $T^{\ast }\mathcal{M}$ (these run from $%
1,\ldots ,2n$).

We start with the phase space of a single classical particle confined to a
general manifold $\left( \mathcal{M},g\right) $. \ The objects needed are
the phase space, $T^{\ast }\mathcal{M}$ which is the cotangent bundle of $%
\mathcal{M}$, an affine connection on the phase space $D$ and the symplectic
form $\omega $ of $T^{\ast }\mathcal{M}$.

A phase-space connection's action on all functions $f\left( x,p\right) \in
T^{\ast }\mathcal{M}$ and a basis of covectors $\Theta ^{A}\in T^{\ast
}T^{\ast }\mathcal{M}$ are:%
\begin{equation*}
Df=df=\frac{\partial f}{\partial x^{a}}dx^{a}+\frac{\partial f}{\partial
p_{a}}dp_{a}
\end{equation*}%
\begin{equation*}
D\otimes \Theta ^{A}=\Gamma _{~B}^{A}\otimes \Theta ^{B}=\Gamma
_{~BC}^{A}\Theta ^{C}\otimes \Theta ^{B}
\end{equation*}%
in such a way as to preserve the symplectic form $\omega =dp_{a}\wedge
dx^{a} $ on $T^{\ast }\mathbb{S}^{2}$ ($D\otimes \omega =0$) where $D=\Theta
^{C}D_{C}$, $D_{C}\Theta ^{A}=\Gamma _{~BC}^{A}\Theta ^{B}$ and $\Gamma
_{~BC}^{A}$ is the Christoffel symbol in this basis.

Additionally we impose that $D$ be torsion-free ($D^{2}f=0$) and that it
corresponds to the Levi-Civita connection on $\mathcal{M}$ when it acts on
functions of $x$ and $dx$. Of course we extend to vectors and higher tensors
by the Leibnitz rule.

In the specific case of $\mathbb{S}^{2}$ ($T^{\ast }\mathbb{S}^{2}$) we
employ the convention that the lower/upper-case indices be of the embedding
space $\mathbb{E}^{3}$ ($T^{\ast }\mathbb{E}^{3}$) running from $1,2,3$ ($%
1,\ldots ,6$) instead of $1,2$ ($1,\ldots ,4$). We note before continuing
that the calculation of the Fedosov observables is \underline{inherently}
two space-time dimensional. The third coordinate is merely for convenience.
We see this fact manifest itself by the two conditions (e.g. $\underline{x}%
\cdot \underline{x}=1$ and $\underline{x}\cdot \underline{p}=0$) on the
three coordinates every step of the way.

The natural objects and quantities on $T^{\ast }\mathbb{S}^{2}$ are:

\begin{itemize}
\item The induced $\mathbb{S}^{2}$ metric $g$ by the $\mathbb{E}^{3}$
embedding metric $\delta $.

\item The induced $T^{\ast }\mathbb{S}^{2}$ symplectic form $\omega $\ by
the $T^{\ast }\mathbb{E}^{3}$ embedding symplectic form.

\item Also the equations defining $T^{\ast }\mathbb{S}^{2}$ inside of $%
T^{\ast }\mathbb{E}^{3},$ $\underline{x}\cdot \underline{x}=\delta
_{ab}x^{a}x^{b}=1$ and $\underline{x}\cdot \underline{p}=x^{a}p_{a}=0$.

\item A torsion-free phase-space connection $D=\Theta ^{A}D_{A}$ on $T^{\ast
}\mathbb{S}^{2}$ that preserves all of the above conditions along with the
symplectic form $\omega $ and there subsequent derivatives. In other words:%
\begin{equation*}
D^{l}\otimes g=D^{l}\otimes \omega =D^{l}\left( \delta
_{ab}x^{a}x^{b}\right) =D^{l}\left( x^{a}p_{a}\right) =0
\end{equation*}
\end{itemize}

for all positive integers $l$\ where $g=g_{ab}dx^{a}\vee dx^{b},~\omega
=\omega _{AB}\Theta ^{A}\wedge \Theta ^{B}$ , where $\Theta ^{A}$ is basis
of forms and $\vee ,\wedge $ are the symmetric, antisymmetric tensor
products respectively that we will omit because it will be clear when we
mean the one or the other.

We define a basis of covectors or forms by:%
\begin{equation*}
\Theta ^{A}=\left( \theta ^{a},\alpha _{a}\right)
\end{equation*}%
where the $\theta $'s are the first three $\Theta $'s and the $\alpha $'s
are the last three $\Theta $'s. $\theta $ and $\alpha $ are defined to be:%
\begin{equation*}
\underline{\alpha }:=\underline{x}\times d\underline{p}
\end{equation*}%
\begin{equation*}
\underline{\theta }:=\underline{x}\times d\underline{x}
\end{equation*}

The metric on $\mathbb{S}^{2}$ is: 
\begin{equation*}
g=\underline{\theta }\cdot \underline{\theta }
\end{equation*}%
The phase-space connection we use for $T^{\ast }\mathbb{S}^{2}$ is:%
\begin{equation*}
D\underline{x}:=d\underline{x}=\underline{\theta }\times \underline{x}
\end{equation*}%
\begin{equation*}
D\underline{p}:=d\underline{p}=\underline{\alpha }\times \underline{x}-%
\underline{p}\times \underline{\theta }
\end{equation*}%
\begin{equation}
D\otimes \underline{\theta }=\underline{\theta }\otimes _{\times }\underline{%
\theta }  \tag{$D\theta $}
\end{equation}%
\begin{equation}
D\otimes \underline{\alpha }=\underline{\theta }\otimes _{\times }\underline{%
\alpha }-\frac{2}{3}\left( \underline{\theta }\times \underline{x}\right)
\otimes \left( \underline{p}\cdot \underline{\theta }\right) +\frac{1}{3}%
\left( \underline{p}\cdot \underline{\theta }\right) \otimes \left( 
\underline{\theta }\times \underline{x}\right)  \tag{$D\alpha $}
\end{equation}%
And its corresponding curvature:%
\begin{equation*}
D^{2}\underline{x}:=0
\end{equation*}%
\begin{equation*}
D^{2}\underline{p}:=0
\end{equation*}%
\begin{equation}
D^{2}\otimes \underline{\theta }=\tilde{\omega}\otimes \left( \underline{x}%
\times \underline{\theta }\right)  \tag{$D^{2}\theta $}
\end{equation}%
\begin{equation}
D^{2}\otimes \underline{\alpha }=\tilde{\omega}\otimes \left( \underline{x}%
\times \underline{\alpha }\right) +\frac{1}{3}\left( \underline{\alpha }%
\left( \underline{\theta }\otimes _{\cdot }\underline{\theta }\right) -%
\underline{\theta }\left( \underline{\alpha }\otimes _{\cdot }\underline{%
\theta }\right) -2\omega \otimes \underline{\theta }\right) 
\tag{$D^{2}\alpha $}
\end{equation}

\section{Introducing the $\hat{y}$'s}

Following Fedosov, we are going to introduce some machinery namely the
operators $\hat{y}$'s to calculate the observables on general manifold $%
\mathcal{M}$. However, unlike Fedosov who defines these $\hat{y}$'s as
covectors equipped with a Moyal-like product between them we choose a
different starting point. We define the $\hat{y}$'s at fixed point to be a
Heisenberg algebra $\left[ \hat{y}^{A},\hat{y}^{B}\right] =i\hbar \omega
^{AB}$ where $\omega ^{AB}$ is the inverse of $\omega _{AB}$\ with $\omega
^{AB}\omega _{BC}=\delta _{C}^{A}$. More explicitly $\hat{y}$'s are huge
(infinite dimensional) matrices that act on a Hilbert space:%
\begin{equation*}
\hat{y}^{A}=\left( 
\begin{array}{ccc}
y_{11}^{A}\left( x,p\right) & y_{12}^{A}\left( x,p\right) & \cdots \\ 
y_{21}^{A}\left( x,p\right) & y_{22}^{A}\left( x,p\right) & \cdots \\ 
\vdots & \vdots & \ddots%
\end{array}%
\right)
\end{equation*}%
where for each $A$, $j$, and $k$ $y_{jk}^{A}\in C^{\infty }\left( T^{\ast }%
\mathcal{M}\right) $.

To make a connection with a more familiar form of the Heisenberg algebra we
use Darboux's theorem. Darboux's theorem says that in the neighborhood of
each point of $q\in T^{\ast }\mathcal{M}$ there exist $2n$ local coordinates 
$\left( \tilde{x}^{1},\ldots ,\tilde{x}^{n},\tilde{p}_{1},\ldots ,\tilde{p}%
_{n}\right) $\footnote{%
Note that these $2n$ coordinates and are different from the $2n+2$ embedding
coordinates $\left( x^{\mu },p_{\mu }\right) $.}, called canonical or
Darboux coordinates, such that the symplectic form $\omega $ may be written
by means of these coordinates as $\omega =d\tilde{p}_{1}d\tilde{x}%
^{1}+\cdots +d\tilde{p}_{n}d\tilde{x}^{n}$. Thus in this coordinate system
at $q$ the $\hat{y}$'s are expressed as $2n$\ operators $\left( \tilde{s}%
^{1},\ldots ,\tilde{s}^{n},\tilde{k}_{1},\ldots ,\tilde{k}_{n}\right) $
which have the commutators $\left[ \tilde{s}^{i},\tilde{s}^{j}\right] =\left[
\tilde{k}_{i},\tilde{k}_{j}\right] =0,~\left[ \tilde{s}^{i},\tilde{k}_{j}%
\right] =i\hbar \delta _{j}^{i}$ where $i$ and $j$ run from $1$ through $2n$%
. And so at each point the $\hat{y}$'s establish a Heisenberg algebra which
acts on a Hilbert space.

\textbf{Important Note:} Fedosov actually begins with the $\hat{y}$'s as
being an arbitrary basis of ordinary covectors with a Moyal-like product
between themselves.$\left[ \text{\hyperlink{1}{1}}\right] $ We take the
point of view that the specific form of the product is irrelevant. All that
matters is that we have an algebra with same commutation relations and the
action of the connection is same on the $\hat{y}$'s.

\underline{Defining Properties of $\hat{y}$:}%
\begin{equation*}
\left[ \hat{y}^{A},\hat{y}^{B}\right] =i\hbar \omega ^{AB}
\end{equation*}%
\begin{equation*}
D\hat{y}^{A}=\Gamma _{~B}^{A}\hat{y}^{B}=\Gamma _{~BC}^{A}\Theta ^{C}\hat{y}%
^{B}~~~,~~~\Theta ^{A}=\left( \theta ^{a},\alpha _{a}\right)
\end{equation*}%
The $\hat{y}$'s commute with the set of quantities $\left\{
x,p,dx,dp,g,\omega ,\hbar ,i\right\} $ where $i$ is the complex unit.

\textbf{Note:} The action of the phase-space connection on $\hat{y}$ is the
same as the one on $\Theta $ ($D\otimes \Theta ^{A}=\Gamma _{~BC}^{A}\Theta
^{C}\otimes \Theta ^{B}$) and so we regard it as a basis of operator or
matrix-valued covectors.\footnote{%
One may be tempted to quantize the manifold by mapping $\left(
x^{1},x^{2},x^{3},p_{1},p_{2},p_{3}\right) $ to the matrices $\left( \hat{y}%
^{1},\hat{y}^{2},\hat{y}^{3},\hat{y}^{4},\hat{y}^{5},\hat{y}^{6}\right) $,
but we want a coordinate independent formalism and, in general, this is not
coordinate independent.} This tells us how to parallel transport the
Heisenberg algebra (the $\hat{y}$'s) at one point to the Heisenberg algebra
of every other point in a consistent way.

\underline{Introducing terminology:}

In this paper when we say $f$ is a function/form we define it to be a
complex Taylor series in its variables\footnote{%
The set of all of these type of functions is sometimes called the enveloping
algebra of its arguments.}. Explicitly:%
\begin{equation}
f\left( u,\ldots ,v\right) =\sum_{l,j\text{'s}}f_{j_{1}\cdots
j_{l}}u^{j_{1}}\cdots v^{j_{l}}\text{ \ \ (}j\text{'s are powers not indices)%
}  \notag
\end{equation}%
where $f_{j_{1}\cdots j_{l}}$\ are constants while $u$ and $v$ could be any
of the set $\left\{ x,p,dx,dp,\omega ,\hbar ,i\right\} $.

So if $f$ is a function/form of some subset or all of the quantities $%
x,p,dx,dp,\omega ,\hbar $ and $i$ it then commutes with the $\hat{y}$'s and
will be called a complex-valued function/form. On the contrary a
matrix-valued function/form is a complex Taylor series in $\hat{y}$ and
possibly some subset or all of the quantities $x,p,dx,dp,\omega ,\hbar $ and 
$i$.

So if $f\left( x,p,dx,dp,\omega ,\hbar ,i\right) $ is a complex-valued
function/form it then commutes with the $\hat{y}$'s. More explicitly with
the matrix indices written:%
\begin{equation*}
\left( \hat{y}^{A}\hat{y}^{B}\right) _{jk}=\Sigma _{l}\hat{y}_{jl}^{A}\hat{y}%
_{lk}^{B}
\end{equation*}%
\begin{equation*}
\left( \left[ \hat{y}^{A},f\right] \right) _{jk}:=\hat{y}_{jk}^{A}f-f\hat{y}%
_{jk}^{A}=0
\end{equation*}%
On the contrary a matrix-valued function/form does not. From now on we will
not write the matrix indices explicitly.

\underline{\textbf{The End Goal:}}

The idea for Fedosov's introduction of the $\hat{y}$'s is to associate to
each $f\left( x,p\right) \in C^{\infty }\left( T^{\ast }\mathcal{M}\right) $
a unique observable $\hat{f}\left( x,p,\hat{y}\right) $:%
\begin{equation}
\hat{f}\left( x,p,\hat{y}\right) =\sum_{l}f_{A_{1}\cdots A_{l}}\hat{y}%
^{A_{1}}\cdots \hat{y}^{A_{l}}  \tag{$\hat{f}$}  \label{fhat}
\end{equation}%
where $f_{A_{1}\cdots A_{l}}$ are some unknown functions of $x$ and $p$\ to
be determined.

\textbf{Important Note:} Most of the rest of the sections will be dedicated
to finding a solution for $\hat{f}$ (i.e. the coefficients functions $%
f_{A_{1}\cdots A_{l}}$) for each $f\left( x,p\right) \in C^{\infty }\left(
T^{\ast }\mathbb{S}^{2}\right) $ up to some "reasonable" ambiguity
(discussed in sections 4.1 and 5).

\subsection{$T^{\ast }\mathbb{S}^{2}$ Explicitly}

Specifically for $T^{\ast }\mathbb{S}^{2}$ we have the induced symplectic
form $\omega $ of $T^{\ast }%
\mathbb{R}
^{3}$ onto $T^{\ast }\mathbb{S}^{2}$ being: 
\begin{equation*}
\omega =\underline{\alpha }\cdot \underline{\theta }=\left( \delta
_{b}^{a}-x^{a}x_{b}\right) \alpha _{a}\theta ^{b}
\end{equation*}%
We make the convention\footnote{%
Note that the indices go from $1$ to $2n+2$ and are different from the $2n$
operators defined above by $\left( \tilde{s}^{1},\ldots ,\tilde{s}^{n},%
\tilde{k}_{1},\ldots ,\tilde{k}_{n}\right) $. The difference between them is
the same as the difference between the embedding coordinates $\left(
x^{1},\ldots ,x^{n+1},p_{1},\ldots ,p_{n+1}\right) $ and $\left( \tilde{x}%
^{1},\ldots ,\tilde{x}^{n},\tilde{p}_{1},\ldots ,\tilde{p}_{n}\right) $.}:%
\begin{equation*}
\hat{y}^{A}=\left( s^{a},k_{a}\right)
\end{equation*}%
where the $s$'s are the first three $\hat{y}$'s and the $k$'s are the last
three $\hat{y}$'s. Using the above formulas we then write the commutation
relations:%
\begin{equation*}
\left[ s^{a},s^{b}\right] =0=\left[ k_{a},k_{b}\right] \ ,\ \left[
s^{a},k_{b}\right] =i\hbar \left( \delta _{b}^{a}-x^{a}x_{b}\right)
\end{equation*}

We may assume w.l.o.g. that $\underline{x}\cdot \underline{s}=\underline{x}%
\cdot \underline{k}=0$ because we observe that the only part of $s$ and $k$
that affect the commutators are the parts that are perpendicular to $x$. The
irrelevance of the part of $s$ and $k$ parallel to $x$ stems from the above
relations because $\left[ x_{a}s^{a},k_{b}\right] =0$ and $\left[
s^{a},k_{b}x^{b}\right] =0$ and so we could always subtract off the part of $%
s$ and $k$ parallel to $x$ and get the same commutators. Since $\underline{x}%
\cdot \underline{s}=\underline{x}\cdot \underline{k}=0$ we have four
independent operators which is required since (one for each direction on $%
T^{\ast }\mathbb{S}^{2}$).

The action of the connection and curvature acting on $\underline{s}$ \& $%
\underline{k}$ is written down directly from the equations $\left( D\theta
\right) ,~\left( D\alpha \right) ,~\left( D^{2}\theta \right) ,$ and $\left(
D^{2}\alpha \right) $:%
\begin{equation*}
D\underline{s}=\underline{\theta }\times \underline{s}
\end{equation*}%
\begin{equation*}
D\underline{k}=\underline{\theta }\times \underline{k}-\frac{2}{3}\underline{%
\theta }\times \underline{x}\left( \underline{p}\cdot \underline{s}\right) +%
\frac{1}{3}\left( \underline{p}\cdot \underline{\theta }\right) \left( 
\underline{s}\times \underline{x}\right)
\end{equation*}%
\begin{equation*}
D^{2}\underline{s}=\tilde{\omega}\left( \underline{x}\times \underline{s}%
\right)
\end{equation*}%
\begin{equation*}
D^{2}\underline{k}=\tilde{\omega}\left( \underline{x}\times \underline{k}%
\right) +\frac{1}{3}\left( \underline{\alpha }\left( \underline{s}\cdot 
\underline{\theta }\right) +\left( \underline{s}\cdot \underline{\alpha }%
\right) \underline{\theta }-2\omega \underline{s}\right)
\end{equation*}

\section{Constructing the global derivation $\hat{D}$}

Following Fedosov, we now introduce a global derivation as a matrix
commutator $\hat{D}=\left[ \hat{Q},\cdot \right] $ which is central to
constructing the coefficients $f_{A_{1}\cdots A_{l}}$ in equation $\left( 
\text{\hyperref[fhat]{$\hat{f}$}}\right) $\ for each $f\left( x,p\right) \in
C^{\infty }\left( T^{\ast }\mathcal{M}\right) $. One possible physical
motivation for $\hat{D}$ is that in the next section we will require that
all observables $\hat{f}$ must satisfy the equation $\left( D-\hat{D}\right) 
\hat{f}\left( x,p,\hat{y}\right) =0$. We see that on $\hat{f}$ $\hat{D}$ is
an infinitesimal translation matrix operator equivalent to $D$. We then
reason that matrix operators corresponding to infinitesimal translations on
the cotangent bundle should exist i.e. $\hat{D}$. The reason that we require
that they must exist is because we are constructing the set of \textit{all}
physical matrix operators on states and certainly infinitesimal translations
are in this set. If this reasoning is correct then the equation $\left( D-%
\hat{D}\right) \hat{f}=0$ must be satisfied for all observables $\hat{f}$.
Also the case of $T^{\ast }%
\mathbb{R}
^{n}$ may provide some insight since it is the overlap of this formalism and
quantum mechanics using the Moyal $\ast $ (see in \hyperlink{D}{Appendix D}
for the example of $T^{\ast }%
\mathbb{R}
^{n}$).

Define the derivation $\hat{D}$ by the graded commutator\footnote{%
Graded commutators have the property that $\left[ \hat{Q}_{A}\Theta ^{A},w%
\right] =\left[ \hat{Q}_{A},w\right] \Theta ^{A}=\left( \hat{Q}_{A}w-w\hat{Q}%
_{A}\right) \Theta ^{A}$ where $w$ is an $l$-form with coefficients $%
w_{A_{1}\cdots A_{l}}$ which are complex-valued functions of the variables $%
x,p$ and $\hat{y}$.}:%
\begin{equation}
\hat{D}=\left[ \hat{Q},\cdot \right] =\left[ \hat{Q}_{A}\Theta ^{A},\cdot %
\right]  \tag{$\hat{D}$}
\end{equation}%
\begin{equation*}
\hat{Q}_{A}=\sum_{l}Q_{AA_{1}\cdots A_{l}}\hat{y}^{A_{1}}\cdots \hat{y}%
^{A_{l}}
\end{equation*}%
where $\Theta ^{A}=\left( \theta ^{a},\alpha _{a}\right) $ and $%
Q_{AA_{1}\cdots A_{l}}$\ are complex-valued functions of $x$ and $p$ that
need to be determined. We reiterate that complex-valued functions are\ not
matrices hence they commute with the $\hat{y}$'s.

Again following Fedosov, we can partially determine the functions $%
Q_{AA_{1}\cdots A_{l}}$ by the mysterious equation\footnote{%
Fedosov adds an additional condition that makes his $\hat{D}$ unique from a
fixed $D$ being $\hat{d}^{-1}r_{0}=0$ where $\hat{d}^{-1}$ is what he calls $%
\delta ^{-1}$ (an operator used in a de Rham decomposition) and $r_{0}$ is
the first term in the recursive solution. We regard this choice as being
artificial and thus omit it from the paper.}:%
\begin{equation}
\left( D-\hat{D}\right) ^{2}\hat{y}^{A}=0  \tag{cond $\hat{D}$}
\label{cond Dhat}
\end{equation}%
The physical motivation for this equation is still unclear and may lurk in
the work of Fedosov. One reason for the above requirement is that in the
next section we want to solve the equation $\left( D-\hat{D}\right) \hat{f}%
=0 $ for $\hat{f}$ and the above is an integrability condition for the
solvability of this equation.

We now let $\hat{Q}$ be the sum of 2 parts the first being the solution in
the case of $T^{\ast }%
\mathbb{R}
^{n}$ (Christoffels$=\Gamma =0$):%
\begin{equation}
\hat{Q}_{A}\Theta ^{A}=\omega _{AB}\hat{y}^{A}\Theta ^{B}+r  \tag{$\hat{Q}$}
\label{Qhat}
\end{equation}%
where:%
\begin{equation*}
r=\sum_{l}r_{AA_{1}\cdots A_{l}}\Theta ^{A}\hat{y}^{A_{1}}\cdots \hat{y}%
^{A_{l}}
\end{equation*}%
and $r_{AA_{1}\cdots A_{l}}$ are complex-valued functions of $x$ and $p$
that need to be determined. In general, we assume that $r$ has terms that
are cubic or higher powers in the $\hat{y}$'s (see \hyperlink{B}{Appendix B}
and Fedosov $\left[ \text{\hyperlink{1}{1}}\right] $ for clarification).

We rewrite the condition $\left( \text{\hyperref[cond Dhat]{cond $\hat{D}$}}%
\right) $ as:%
\begin{equation*}
\left( D-\hat{D}\right) ^{2}\hat{y}^{A}=\left[ \Omega -Dr+\hat{d}r+r^{2},%
\hat{y}^{A}\right] =0
\end{equation*}%
where $\Omega :=\frac{1}{2i\hbar }\omega _{FN}R_{~BCE}^{F}\Theta ^{C}\Theta
^{E}\hat{y}^{N}\hat{y}^{B}$ is the phase-space curvature ($D^{2}\otimes
\Theta ^{A}=R_{~BCE}^{A}\Theta ^{C}\Theta ^{E}\otimes \Theta ^{B}$) as a
commutator and $\hat{d}h=\frac{1}{i\hbar }\left[ \omega _{AB}\hat{y}%
^{A}\Theta ^{B},h\right] $ where $h$ is a matrix-valued function of $%
x,~p,~dx,~dp$ and $\hat{y}$ (see \hyperlink{A}{Appendix A} for the proof).

From now on we let: 
\begin{equation}
\Omega -Dr+\hat{d}r+r^{2}=0  \tag{$r$}  \label{r}
\end{equation}%
and keep it in the back of our minds that we could add something that
commutes with all $\hat{y}$'s to $\Omega -Dr+\hat{d}r+r^{2}$.

\textbf{Important:} To emphasize the importance of this equation the reader
should note that the whole Fedosov $\ast $-formalism hinges on this $r$
existing. We know solutions exists perturbatively in general (Fedosov has
the recursive solution for it $\left[ \text{\hyperlink{1}{1}}\right] \left[
p.144\right] $), however convergence issues still remain unresolved. On a
technical note we have found that solving for $r$ to be the hardest point of
the computation of the Fedosov observables because of the need for the right
ansatz and the nonlinear equation $\left( \hyperref[r]{r}\right) $ above
that it must solve.

Specifically for the case of $T^{\ast }\mathbb{S}^{2}$\ the solution for the
curvature as a commutator $\Omega $ is:%
\begin{equation*}
\Omega :=\frac{1}{3}\left( \left( \underline{s}\cdot \underline{\alpha }%
\right) \left( \underline{s}\cdot \underline{\theta }\right) -s^{2}\omega
\right) +\left( \underline{x}\times \underline{k}\right) \cdot \underline{s}%
\tilde{\omega}
\end{equation*}%
We then verify that it gives the curvature as commutators:%
\begin{equation*}
\left[ \Omega ,\underline{s}\right] =\left[ -\underline{k}\cdot \left( 
\underline{x}\times \underline{s}\right) \tilde{\omega},\underline{s}\right]
=\tilde{\omega}\left( \underline{x}\times \underline{s}\right)
\end{equation*}%
\begin{equation*}
\left[ \Omega ,\underline{k}\right] =\frac{1}{3}\left( \underline{\alpha }%
\left( \underline{s}\cdot \underline{\theta }\right) +\left( \underline{s}%
\cdot \underline{\alpha }\right) \underline{\theta }-2\omega \underline{s}%
\right) +\left( \underline{x}\times \underline{k}\right) \tilde{\omega}
\end{equation*}%
To simplify the calculations we set $i\hbar =1$ which we will eventually put
back in the end.

Fedosov at this point would implement an algorithm to construct $r$
perturbatively, however rather than do this we will make an ansatz for $r$
by exploiting the rotational symmetry of the sphere. This will give us an
exact solution for $r$.\footnote{%
On a technical note: we ran the Fedosov algorithm a few times to help us see
what form the ansatz should take. Also remember that when we require $\Omega
-Dr+\hat{d}r+r^{2}=0$ modulo terms that commute with the $\hat{y}$'s.}

Our ansatz for $r$ is:%
\begin{equation}
r=r_{0}+f\left( s^{2}\right) \underline{z}\cdot \underline{s}\left( 
\underline{x}\times \underline{s}\right) \cdot \underline{\theta }+g\left(
s^{2}\right) \underline{z}\cdot \left( \underline{x}\times \underline{s}%
\right) \underline{s}\cdot \underline{\theta }+h\left( s^{2}\right) 
\underline{s}\cdot \underline{\theta }  \tag{r ansatz}  \label{r ansatz}
\end{equation}%
where $\underline{z}=\underline{p}-\underline{x}\times \underline{k}$ and $%
r_{0}=\frac{1}{3}\left( \left( \underline{k}\cdot \underline{\theta }\right)
s^{2}-\underline{k}\cdot \underline{s}\left( \underline{s}\cdot \underline{%
\theta }\right) \right) $.

We will now state the results of our calculations because the calculations
are just too space consuming and yet at the same time straight forward.\
Given the formulas for $r$ and $\Omega $ and performing lengthy calculations
eventually we get:%
\begin{equation*}
Dr=\left( \frac{1}{9}-\frac{2g}{3}+\frac{f}{3}\right) s^{2}\underline{p}%
\cdot \underline{s}\tilde{\omega}+f\underline{\alpha }\cdot \left( 
\underline{x}\times \underline{s}\right) \left( \underline{x}\times 
\underline{s}\right) \cdot \underline{\theta }-g\left( \underline{s}\cdot 
\underline{\alpha }\right) \underline{s}\cdot \underline{\theta }
\end{equation*}%
\begin{equation*}
\hat{d}r=-\Omega +\left( 2f^{\prime }s^{2}+3f+g\right) \underline{z}\cdot 
\underline{s}\tilde{\omega}-g\left( \underline{s}\cdot \underline{\alpha }%
\right) \underline{s}\cdot \underline{\theta }+f\underline{\alpha }\cdot
\left( \underline{x}\times \underline{s}\right) \left( \underline{x}\times 
\underline{s}\right) \cdot \underline{\theta }
\end{equation*}%
\begin{equation*}
r^{2}=\left( \frac{1}{9}-\frac{2g}{3}+\frac{f}{3}\right) s^{2}\underline{p}%
\cdot \underline{s}\tilde{\omega}+\left( 2gf^{\prime }s^{2}+gf-f^{2}-\frac{2f%
}{3}+\frac{g}{3}-\frac{1}{9}\right) s^{2}\underline{z}\cdot \underline{s}%
\tilde{\omega}
\end{equation*}%
$\allowbreak $where $f^{\prime }=\frac{\partial f}{\partial \left(
s^{2}\right) }$ for all functions.

Putting these into the equation $\left( \hyperref[r]{r}\right) $ we obtain a
condition for $g$:%
\begin{equation*}
g=\frac{s^{2}\left( \left( f+\frac{1}{3}\right) ^{2}-2f^{\prime }\right)
\allowbreak -3f}{s^{2}\left( \left( f+\frac{1}{3}\right) +2s^{2}f^{\prime
}\right) +1}
\end{equation*}%
while $f$ and $h$ are left arbitrary as long as $g$ is well-defined. This is
a necessary and sufficient condition for the equation $\left( \hyperref[r]{r}%
\right) $\ to hold.

We note that $f=-\frac{1}{3},g=1$ and $f=-\frac{1}{12},g=\frac{1}{4}$ are
the only solutions where $f$ and $g$ are constant. We will choose to work
with the $f=-\frac{1}{3},~g=1,~h=0$ solution from now on. We choose this
solution for the sake of clarity because it turns out to be the easiest to
use in the next few sections. However the reader should note that we
calculated the commutators for the general solutions for $g,~f$ and $h$ and
obtained the same result for all of them. See section 6 for the exact result
of the commutators for the particular solution $f=-\frac{1}{3},~g=1,~h=0$
(and hence the solution for the general solutions for $g,~f$ and $h$).

The solution for $r$ for $f=-\frac{1}{3},~g=1,~h=0$ is:%
\begin{equation}
r=-\frac{1}{3}\left( \underline{p}\cdot \underline{s}\right) \left( \left( 
\underline{x}\times \underline{s}\right) \cdot \underline{\theta }\right) +%
\underline{z}\cdot \left( \underline{x}\times \underline{s}\right) 
\underline{s}\cdot \underline{\theta }  \tag{r soln}  \label{r soln}
\end{equation}

\subsection{Ambiguities in $r$}

It is worthwhile to note that the condition $\left( \text{\text{\hyperref[cond Dhat%
]{cond $\hat{D}$}}}\right) $ does not uniquely define $\hat{D}$ given a
fixed $D$.\footnote{%
Fedosov adds an additional condition that makes his $\hat{D}$ unique from a
fixed $D$ being $\hat{d}^{-1}r_{0}=0$ where $\hat{d}^{-1}$ is what he calls $%
\delta ^{-1}$ (an operator used in a de Rham decomposition) and $r_{0}$ is
the first term in the recursive solution. We regard this choice as being
artificial and thus omit it from the paper.} It appears however that the
most of the ambiguities in constructing $\hat{D}$ when given a fixed phase
space connection $D$ can be absorbed by a basis change (in other words a
gauge transformation). It is easy to see this in a Darboux chart because the
connection may be expressed as a commutator:%
\begin{equation*}
\tilde{D}\hat{y}^{A}=\left[ \tilde{Q},\hat{y}^{A}\right]
\end{equation*}%
where $\tilde{D}=D-\hat{D}$, $\tilde{Q}=Q-\hat{Q}$ and $D=\left[ Q,\cdot %
\right] $. The gauge transformation takes the form: 
\begin{equation*}
\hat{y}^{A}\rightarrow \hat{y}_{new}^{A}:=U\hat{y}^{A}U^{-1}\text{ \ \ , \ \ 
}\tilde{D}\hat{y}^{A}\rightarrow \tilde{D}_{new}\hat{y}_{new}^{A}:=\left[ U%
\tilde{Q}U^{-1},U\hat{y}^{A}U^{-1}\right] =U\left( \tilde{D}\hat{y}%
^{A}\right) U^{-1}
\end{equation*}%
where $U$ is some invertible function of the $x$'s, $p$'s and $\hat{y}$'s.
Thus the physical content of this theory is independent of $U$ because the
commutators remain unchanged.

This can be seen as follows:%
\begin{equation*}
r\rightarrow r+r^{\prime }
\end{equation*}%
where $r$ is a solution to the equation $\left( \hyperref[r]{r}\right) $ and 
$r^{\prime }$ is some unknown series:%
\begin{equation*}
r^{\prime }=\sum_{l}r_{AA_{1}\cdots A_{l}}^{\prime }\Theta ^{A}\hat{y}%
^{A_{1}}\cdots \hat{y}^{A_{l}}
\end{equation*}%
Putting $r\rightarrow r+r^{\prime }$ into $\left( \hyperref[r]{r}\right) $
we obtain:

\begin{equation*}
\Omega -D\left( r+r^{\prime }\right) +\left[ \underline{s}\cdot \underline{%
\alpha }-\underline{k}\cdot \underline{\theta },\left( r+r^{\prime }\right) %
\right] +\left( r+r^{\prime }\right) ^{2}=0
\end{equation*}%
modulo the equation $\left( \hyperref[r]{r}\right) $\ to get:%
\begin{equation*}
-Dr^{\prime }+\left[ \underline{s}\cdot \underline{\alpha }-\underline{k}%
\cdot \underline{\theta },r^{\prime }\right] +\left( r^{\prime }\right) ^{2}+%
\left[ r,r^{\prime }\right] =0
\end{equation*}%
\begin{equation*}
\implies \tilde{D}r^{\prime }-\left( r^{\prime }\right) ^{2}=0
\end{equation*}%
This tells us that if $r^{\prime }$ is of the form:%
\begin{equation*}
r^{\prime }=\left( \tilde{D}U\right) U^{-1}
\end{equation*}%
for any $U$ which corresponds to a gauge transformation in the enveloping
algebra then the resulting $r_{new}=r+r^{\prime }$ will solve equation $%
\left( \hyperref[r]{r}\right) $. \ In other words once we have one solution
we have actually have huge class of equivalent solutions. We suspect this
class of equivalent solutions are all of the solutions for a
simply-connected manifold.

\textbf{Note:} There is another source of ambiguity namely the ambiguity in
the phase-space connection $D$.\ Given a connection $D$ we may add to it a
tensor $\Delta _{\text{ }BC}^{A}$ where if we lower by $\Delta _{ABC}=\omega
_{AE}\Delta _{\text{ \ }BC}^{E}$ it is symmetric in all three indices. The
new connection still preserves the symplectic form $\omega $. Our curvature
becomes:%
\begin{equation*}
\left( D+\Delta \right) ^{2}=D^{2}+D\left( \Delta \right) +\Delta ^{2}
\end{equation*}%
It is unclear what this ambiguity means so we will leave it for a future
discussion.

\section{Computing $\hat{x}$ and $\hat{p}$}

At this point in Fedosov's algorithm we have all the tools in place to
associate an observable $\hat{f}$ to every $f\in C^{\infty }\left( T^{\ast }%
\mathcal{M}\right) $. Following Fedosov we require that every observable $%
\hat{f}\left( x,p,\hat{y}\right) $ must satisfy the equation:%
\begin{equation}
\left( D-\hat{D}\right) \hat{f}\left( x,p,\hat{y}\right) =0  \notag
\end{equation}%
where $f_{A_{1}\cdots A_{l}}$ are some unknown functions of $x$ and $p$\
such that: 
\begin{equation*}
\ell o\left( \hat{f}\left( x,p,\hat{y}\right) \right) =f\left( x,p\right)
\end{equation*}%
$\ell o$ (short for leading order in $\hat{y}$ and $\hbar $) picks out the
term which has no $\hat{y}$'s and no $\hbar $'s in it. Explicitly:%
\begin{equation}
\hat{f}\left( x,p,\hat{y}\right) =f\left( x,p\right) +\mathcal{O}\left( \hat{%
y},\hbar \right)  \notag
\end{equation}

where $f$ has no $\hbar $'s in it.

And so the condition to solve (we believe unique up to unitary
transformations) for an observable $\hat{f}$ for every $f\in C^{\infty
}\left( T^{\ast }\mathcal{M}\right) $ is:%
\begin{equation}
\left( D-\hat{D}\right) \hat{f}\left( x,p,\hat{y}\right) =0~~~,~~~\ell
o\left( \hat{f}\left( x,p,\hat{y}\right) \right) =f\left( x,p\right) 
\tag{cond $\hat{f}$}  \label{cond fhat}
\end{equation}

If we have determined our $D$ and $\hat{D}$ we can find solutions for the
operators $\hat{x}^{a}$ and $\hat{p}_{a}$ (i.e. their coefficients $%
b_{A_{1}\cdots A_{l}}^{a}\,$\ and $c_{aA_{1}\cdots A_{l}}$):%
\begin{equation}
\hat{x}^{a}=\sum_{l}b_{A_{1}\cdots A_{l}}^{a}\hat{y}^{A_{1}}\cdots \hat{y}%
^{A_{l}}  \tag{$\hat{x}$}  \label{xhat}
\end{equation}%
\begin{equation}
\hat{p}_{a}=\sum_{l}c_{aA_{1}\cdots A_{l}}\hat{y}^{A_{1}}\cdots \hat{y}%
^{A_{l}}  \tag{$\hat{p}$}  \label{phat}
\end{equation}%
where $b_{A_{1}\cdots A_{l}}^{a}\,$\ and $c_{aA_{1}\cdots A_{l}}$ are
complex-valued functions of $x$ and $p$ (which are the coefficients $%
f_{A_{1}\cdots A_{l}}$ in equation $\left( \text{\hyperref[fhat]{$\hat{f}$}}%
\right) $ where the first terms in the series is $f=b^{a}=x^{a}$ or $%
f=c_{a}=p_{a}\,$\ respectively) and will be determined by the equations:%
\begin{equation}
\left( D-\hat{D}\right) \hat{x}^{a}=0~~~~,~~\ ~\ell o\left( \hat{x}%
^{a}\right) =x^{a}  \tag{cond $\hat{x}$}  \label{cond xhat}
\end{equation}%
\begin{equation}
\left( D-\hat{D}\right) \hat{p}_{a}=0~~~,~~~\ell o\left( \hat{p}_{a}\right)
=p_{a}  \tag{cond $\hat{p}$}  \label{cond phat}
\end{equation}%
Again see the example in \hyperlink{D}{Appendix D} for solutions to $\hat{x}$
and $\hat{p}$ in the case of $T^{\ast }%
\mathbb{R}
^{n}$ where $D=d$.

If we invert the equations $\left( \text{\hyperref[xhat]{$\hat{x}$}}\right) $
and $\left( \text{\hyperref[phat]{$\hat{p}$}}\right) $ once we have solved
for the coefficients $b_{A_{1}\cdots A_{l}}^{a}$ and $c_{A_{1}\cdots
A_{l}}^{a}$ to get $\hat{y}$ as matrix-valued function of $x,p,\hat{x}$ and $%
\hat{p}$ (i.e. $\hat{y}^{A}=\hat{y}^{A}\left( x,p,\hat{x},\hat{p}\right) $)
and then substitute it into the equation for an arbitrary observable $\left( 
\text{\hyperref[fhat]{$\hat{f}$}}\right) $ and get:%
\begin{equation}
\hat{f}\left( \hat{x},\hat{p}\right) =\sum_{lm}f_{a_{1}\cdots
a_{l}}^{b_{1}\cdots b_{m}}\hat{x}^{a_{1}}\cdots \hat{x}^{a_{l}}\hat{p}%
_{b_{1}}\cdots \hat{p}_{b_{m}}  \tag{$\hat{f}$ soln}  \label{fhat soln}
\end{equation}%
where $f_{a_{1}\cdots a_{l}}^{b_{1}\cdots b_{m}}$ are constant coefficients.%
\footnote{%
To prove this act $D-\hat{D}$ on this equation.}

However, once have our $\hat{x}$ and $\hat{p}$ there is the ambiguity of how
to order each variable when you map a function $f\left( x,p\right) $ to $%
\hat{f}\left( \hat{x},\hat{p}\right) $. For example does the function $%
f\left( x,p\right) =x^{1}p_{1}$ go to $\hat{x}^{1}\hat{p}_{1}$, $\hat{p}_{1}%
\hat{x}^{1}$ or some linear combination of the two? We should expect this in
any well defined quantization procedure because such ordering ambiguities
arise in quantum mechanics. We will, for now, regard the ordering of each $%
\hat{f}$ to be undetermined.\footnote{%
Fedosov chooses Weyl ordering.}

\subsection{$T^{\ast }\mathbb{S}^{2}$ Explicitly}

Fedosov at this point would implement an algorithm to construct $\hat{x}$
and $\hat{p}$ perturbatively$\left[ \text{\hyperlink{1}{1}}\right] \left[
p.146\right] $ for our specific case of $T^{\ast }\mathbb{S}^{2}$. We
instead try to find exact solutions to them.\footnote{%
We, again, ran the Fedosov algorithm a few times to help us see what for the
ansatz should take.} Specifically for the case of $T^{\ast }\mathbb{S}^{2}$
we have the ansatz for both $\hat{x}$ and $\hat{p}$ as:%
\begin{equation*}
\underline{\hat{x}}=v\left( s^{2}\right) \underline{x}+w\left( s^{2}\right) 
\underline{x}\times \underline{s}+y\left( s^{2}\right) \underline{s}
\end{equation*}%
\begin{equation*}
\underline{\hat{p}}=\left( \underline{z}\cdot \underline{s}t\left(
s^{2}\right) +\underline{z}\cdot \left( \underline{x}\times \underline{s}%
\right) q\left( s^{2}\right) \right) \underline{x}+\underline{z}n\left(
s^{2}\right) +\underline{z}\times \underline{x}u\left( s^{2}\right)
\end{equation*}%
with some functions $v,~w,~y,~t,~q,~n$ and $u$ to be determined and the
requirements that $\ell o\left( \underline{\hat{x}}\right) =\underline{x}$
and $\ell o\left( \underline{\hat{p}}\right) =\underline{p}$.

The conditions $\left( \text{\hyperref[cond xhat]{cond $\hat{x}$}}\right) $
and $\left( \text{\hyperref[cond phat]{cond $\hat{p}$}}\right) $\ become the
following equations:%
\begin{eqnarray*}
0 &=&\left( D-\hat{D}\right) \underline{\hat{x}}=\left( \left( -2v^{\prime
}\left( s^{2}+1\right) +w\right) \left( \underline{s}\cdot \underline{\theta 
}\right) -y\left( \underline{x}\times \underline{s}\right) \cdot \underline{%
\theta }\right) \underline{x} \\
&&+\left( \left( -\frac{v}{s^{2}}-2w^{\prime }\left( s^{2}+1\right) -w\left(
1+\frac{1}{s^{2}}\right) \right) \left( \underline{s}\cdot \underline{\theta 
}\right) -y\frac{1}{s^{2}}\left( \underline{x}\times \underline{s}\right)
\cdot \underline{\theta }\right) \underline{x}\times \underline{s} \\
&&+\left( \left( \frac{v}{s^{2}}+w\frac{1}{s^{2}}\right) \left( \underline{x}%
\times \underline{s}\right) \cdot \underline{\theta }+\left( -2y^{\prime
}\left( s^{2}+1\right) -y\left( 1+\frac{1}{s^{2}}\right) \right) \left( 
\underline{s}\cdot \underline{\theta }\right) \right) \underline{s}
\end{eqnarray*}%
and:%
\begin{eqnarray*}
0 &=&\left( D-\hat{D}\right) \underline{\hat{p}}=\left( 
\begin{array}{c}
\left( 
\begin{array}{c}
-2\underline{z}\cdot \underline{s}t^{\prime }\left( s^{2}+1\right) -\left( 
\underline{z}\cdot \underline{s}\right) \frac{1}{s^{2}}t-2\underline{z}\cdot
\left( \underline{x}\times \underline{s}\right) q^{\prime }\left(
s^{2}+1\right) \\ 
+\underline{z}\cdot \left( \underline{x}\times \underline{s}\right) \left( 1-%
\frac{1}{s^{2}}\right) q-\left( \underline{z}\cdot \underline{s}\right) 
\frac{1}{s^{2}}u+\underline{z}\cdot \left( \underline{x}\times \underline{s}%
\right) \frac{1}{s^{2}}n%
\end{array}%
\right) \left( \underline{s}\cdot \underline{\theta }\right) \\ 
\left( 
\begin{array}{c}
-\left( \underline{z}\cdot \left( \underline{x}\times \underline{s}\right)
\left( 1+\frac{1}{s^{2}}\right) \right) t+\left( \underline{z}\cdot 
\underline{s}\right) \frac{1}{s^{2}}q \\ 
+\underline{x}\cdot \left( \underline{z}\times \underline{s}\right) \frac{1}{%
s^{2}}u-\left( \underline{z}\cdot \underline{s}\right) \frac{1}{s^{2}}n%
\end{array}%
\right) \left( \underline{x}\times \underline{s}\right) \cdot \underline{%
\theta }%
\end{array}%
\right) \underline{x} \\
&&+\left( 
\begin{array}{c}
\left( 
\begin{array}{c}
-\underline{z}\cdot \underline{s}t-\underline{z}\cdot \left( \underline{x}%
\times \underline{s}\right) q+2\underline{z}\cdot \left( \underline{x}\times 
\underline{s}\right) n \\ 
-\left( \underline{z}\cdot \underline{s}\right) u+2\left( \underline{z}\cdot 
\underline{s}\right) \left( s^{2}+1\right) u^{\prime }-2\left( \underline{z}%
\times \underline{x}\right) \cdot \underline{s}\left( s^{2}+1\right)
n^{\prime }%
\end{array}%
\right) \left( \underline{s}\cdot \underline{\theta }\right) \\ 
+\underline{z}\cdot \left( \underline{x}\times \underline{s}\right) \left( 
\underline{x}\times \underline{s}\right) \cdot \underline{\theta }u%
\end{array}%
\right) \frac{1}{s^{2}}\underline{x}\times \underline{s} \\
&&+\left( 
\begin{array}{c}
\left( 
\begin{array}{c}
2\underline{z}\cdot \left( \underline{x}\times \underline{s}\right) u+\left( 
\underline{z}\cdot \underline{s}\right) n \\ 
-2\left( \underline{z}\cdot \underline{s}\right) \left( s^{2}+1\right)
n^{\prime }-2\left( \left( \underline{z}\times \underline{x}\right) \cdot 
\underline{s}\right) \left( s^{2}+1\right) u^{\prime }%
\end{array}%
\right) \underline{s}\cdot \underline{\theta } \\ 
+\left( \underline{z}\cdot \underline{s}t+\underline{z}\cdot \left( 
\underline{x}\times \underline{s}\right) q-\underline{z}\cdot \left( 
\underline{x}\times \underline{s}\right) n\right) \left( \underline{x}\times 
\underline{s}\right) \cdot \underline{\theta }%
\end{array}%
\right) \frac{1}{s^{2}}\underline{s}
\end{eqnarray*}

So the conditions that $\tilde{D}\underline{\hat{x}}=0$ and $\tilde{D}%
\underline{\hat{p}}=0$ becomes 6+6 equations because $\left( \underline{s}%
\cdot \underline{\theta }\right) ^{2}=0=\left( \left( \underline{x}\times 
\underline{s}\right) \cdot \underline{\theta }\right) ^{2}$ and $\left( 
\underline{s}\cdot \underline{\theta }\right) \left( \underline{x}\times 
\underline{s}\right) \cdot \underline{\theta }=\tilde{\omega}$ where $\tilde{%
\omega}_{ab}$ is invertible. We then solve the subsequent differential
equations for the functions $v,~w,~y,~t,~q,~n$ and $u$ along with requiring
that they have the correct term with no $\hat{y}$'s ($\ell o\left( 
\underline{\hat{x}}\right) =\underline{x}$ and $\ell o\left( \underline{\hat{%
p}}\right) =\underline{p}$) in the Taylor expansion to obtain the solutions:%
\begin{equation}
\underline{\hat{x}}=\left( \underline{x}-\underline{x}\times \underline{s}%
\right) \left( s^{2}+1\right) ^{-\frac{1}{2}}  \tag{$\hat{x}$ soln}
\end{equation}%
\begin{equation}
\underline{\hat{p}}=\left( \underline{z}\cdot \left( \underline{x}\times 
\underline{s}\right) \underline{x}+\underline{z}\right) \left(
s^{2}+1\right) ^{\frac{1}{2}}  \tag{$\hat{p}$ soln}  \label{phat soln}
\end{equation}%
where $\underline{z}=\underline{p}-\underline{x}\times \underline{k}$ with
the following conditions holding:%
\begin{equation*}
\ell o\left( \underline{\hat{x}}\right) =\underline{x}~~~,~~~\ell o\left( 
\underline{\hat{p}}\right) =\underline{p}
\end{equation*}%
\begin{equation}
\underline{\hat{p}}\cdot \underline{\hat{x}}=\underline{\hat{x}}\cdot 
\underline{\hat{p}}-2i\hbar =0  \tag{$\hat{x}\hat{p}$ conds}
\end{equation}%
We note at this point that there is not much insight looking at these
formulas except for what we get for the commutators in the next
section.\pagebreak 

\section{The Commutators $\left[ \hat{x}^{a},\hat{x}^{b}\right] ,\left[ \hat{%
x}^{a},\hat{p}_{b}\right] $ and $\left[ \hat{p}_{a},\hat{p}_{b}\right] $}

Once we have $\hat{x}^{a}$ and $\hat{p}_{a}$ i.e. the coefficients $%
b_{A_{1}\cdots A_{l}}^{a}$ and $c_{A_{1}\cdots A_{l}}^{a}$\ we work out the
commutation relations $\left[ \hat{x}^{a},\hat{x}^{b}\right] ,\left[ \hat{x}%
^{a},\hat{p}_{b}\right] $ and $\left[ \hat{p}_{a},\hat{p}_{b}\right] $ using
the formulas $\left( \text{\hyperref[xhat]{$\hat{x}$}}\right) $ and $\left( 
\text{\hyperref[phat]{$\hat{p}$}}\right) $ in the previous section in a
brute force calculation. Remember that the $\ast $-commutators is the 
\newline
Poisson bracket on $T^{\ast }\mathcal{M}$ to first order in $\hbar $:%
\begin{equation*}
\left[ \hat{f}\left( \hat{x},\hat{p}\right) ,\hat{g}\left( \hat{x},\hat{p}%
\right) \right] =\hat{h}\left( \hat{x},\hat{p}\right)
\end{equation*}%
\begin{equation}
\left[ f_{\ast }\left( x,p\right) ,g_{\ast }\left( x,p\right) \right] _{\ast
}=h_{\ast }\left( x,p\right) =i\hbar \left\{ f,g\right\} _{\mathcal{M}}+%
\mathcal{O}\left( \hbar ^{2}\right)  \tag{$\ast $-comm}  \label{*-comm}
\end{equation}%
where $\hat{f}$, $\hat{g}$, $\hat{h}$ and $f_{\ast }$, $g_{\ast }$, $h_{\ast
}$ are fuctions defined by:%
\begin{equation*}
\hat{f}\left( \hat{x},\hat{p}\right) =\sum_{lm}f_{ja_{1}\cdots
a_{l}}^{b_{1}\cdots b_{m}}\hbar ^{j}\hat{x}^{a_{1}}\cdots \hat{x}^{a_{l}}%
\hat{p}_{b_{1}}\cdots \hat{p}_{b_{m}}
\end{equation*}%
\begin{equation*}
f_{\ast }\left( x,p\right) =\sum_{lm}f_{ja_{1}\cdots a_{l}}^{b_{1}\cdots
b_{m}}\hbar ^{j}x^{a_{1}}\ast \cdots \ast x^{a_{l}}\ast p_{b_{1}}\ast \cdots
\ast p_{b_{m}}
\end{equation*}%
where $f_{ja_{1}\cdots a_{l}}^{b_{1}\cdots b_{m}}$ are constants.

These two sets, one of all $f_{\ast }$'s $\left\{ f_{\ast }\right\} $ and
one of all $\hat{f}$'s $\left\{ \hat{f}\right\} $ defined above are
isomorphic.\pagebreak 

\subsection{$T^{\ast }\mathbb{S}^{2}$ Explicitly}

In our case of $T^{\ast }\mathbb{S}^{2}$ we find:%
\begin{equation*}
\left[ \hat{x}^{a},\hat{x}^{b}\right] =0
\end{equation*}%
\begin{equation*}
\left[ \hat{x}^{a},\hat{p}_{b}\right] =i\hbar \left( \delta _{b}^{a}-\hat{x}%
^{a}\hat{x}_{b}\right)
\end{equation*}%
\begin{equation*}
\left[ \hat{p}_{a},\hat{p}_{b}\right] =2i\hbar \hat{x}_{[b}\hat{p}_{a]}
\end{equation*}%
\begin{equation*}
\underline{\hat{x}}\cdot \underline{\hat{x}}=1,\text{ }\underline{\hat{p}}%
\cdot \underline{\hat{x}}=\underline{\hat{x}}\cdot \underline{\hat{p}}%
-2i\hbar =0
\end{equation*}

We now define $\underline{\hat{L}}$ because we argue below that it is a more
"natural" momentum:%
\begin{equation*}
\underline{\hat{L}}:=-\underline{\hat{p}}\times \underline{\hat{x}}=%
\underline{\hat{x}}\times \underline{\hat{p}}=\underline{x}\times \underline{%
z}+\left( \underline{z}\cdot \underline{s}\right) \underline{x}-\underline{z}%
\cdot \left( \underline{x}\times \underline{s}\right) \underline{s}
\end{equation*}%
again with the computed conditions:%
\begin{equation*}
\underline{\hat{L}}\cdot \underline{\hat{x}}=\underline{\hat{x}}\cdot 
\underline{\hat{L}}=0~~~,~~~\underline{\hat{x}}\cdot \underline{\hat{x}}=1
\end{equation*}%
\begin{equation*}
\ell o\left( \underline{\hat{L}}\right) =\underline{L}=\underline{x}\times 
\underline{p}
\end{equation*}%
We easily recognize that $\underline{\hat{L}}$ is the more "natural"
variable compared to $\underline{\hat{p}}$. This is because $\underline{\hat{%
p}}\cdot \underline{\hat{x}}=0$ and $\underline{\hat{x}}\cdot \underline{%
\hat{p}}=2i\hbar $ are very "unnatural" conditions since there is no
physical reason why it shouldn't be $\underline{\hat{x}}\cdot \underline{%
\hat{p}}=0$ and $\underline{\hat{p}}\cdot \underline{\hat{x}}=-2i\hbar $. We
could define $\underline{\hat{p}}_{new}=\underline{\hat{p}}+A\underline{\hat{%
x}}$ where $A$ is an arbitrary constant and obtain the same commutators. On
the other hand the symmetry between $\underline{\hat{L}}\cdot \underline{%
\hat{x}}=\underline{\hat{x}}\cdot \underline{\hat{L}}=0$ seems to suggest
that $\underline{\hat{L}}$ should\ be the preferred quantity over $%
\underline{\hat{p}}$. \ In other words the relevant component of $\underline{%
\hat{p}}$ is the one perpendicular to $\underline{\hat{x}}$ which is
precisely what $\underline{\hat{L}}$\ is.

Therefore the part of $\hat{p}$ parallel to $\hat{x}$ is irrelevant:%
\begin{equation}
\underline{\hat{x}}=\left( \underline{x}-\underline{x}\times \underline{s}%
\right) \left( s^{2}+1\right) ^{-\frac{1}{2}}  \tag{$\hat{x}$ soln}
\label{xhat soln}
\end{equation}%
\begin{equation}
\underline{\hat{L}}=\underline{x}\times \underline{z}+\left( \underline{z}%
\cdot \underline{s}\right) \underline{x}-\underline{z}\cdot \left( 
\underline{x}\times \underline{s}\right) \underline{s}  \tag{$\hat{L}$ soln}
\label{Lhat soln}
\end{equation}%
where $\underline{z}=\underline{p}-\underline{x}\times \underline{k}$ with
conditions:%
\begin{equation}
\underline{\hat{L}}\cdot \underline{\hat{x}}=\underline{\hat{x}}\cdot 
\underline{\hat{L}}=0~~~,~~~\underline{\hat{x}}\cdot \underline{\hat{x}}=1 
\tag{$\hat{x}\hat{L}$ conds}  \label{xhatLhat conds}
\end{equation}%
Again we note at this point that there is not much insight looking at these
formulas except for what we get for the commutators in the remainder of this
section.

We compute the commutators:%
\begin{equation}
\left[ \hat{x}^{a},\hat{x}^{b}\right] =0  \tag{$xx$}  \label{xx}
\end{equation}%
\begin{equation}
\left[ \hat{x}^{a},\hat{L}_{b}\right] =i\hbar \varepsilon _{~bc}^{a}\hat{x}%
^{c}  \tag{$xL$}  \label{xL}
\end{equation}%
\begin{equation}
\left[ \hat{L}_{a},\hat{L}_{b}\right] =i\hbar \varepsilon _{~ab}^{c}\hat{L}%
_{c}  \tag{$LL$}  \label{LL}
\end{equation}%
along with:%
\begin{equation}
\underline{\hat{x}}\cdot \underline{\hat{x}}=1,\text{ }\underline{\hat{L}}%
\cdot \underline{\hat{x}}=\underline{\hat{x}}\cdot \underline{\hat{L}}=0 
\tag{cond $xL$}  \label{cond xL}
\end{equation}%
Once we know these relations we know the whole algebra of functions since
the algebra is associative. And thus we are done!

And so in the case of $T^{\ast }\mathbb{S}^{2}$\ a general element $\hat{f}$
(the function $\left( \text{\hyperref[fhat]{$\hat{f}$}}\right) $ we were
looking for and the specific form of the solution $\left( \text{\hyperref[fhat soln%
]{$\hat{f}$ soln}}\right) $)\ in the space of all observables of $\hat{x}$
and $\hat{L}$ is%
\begin{equation*}
\hat{f}\left( \hat{x},\hat{L}\right) =\sum_{lm}f_{a_{1}\cdots
a_{l}}^{b_{1}\cdots b_{m}}\hat{x}^{a_{1}}\cdots \hat{x}^{a_{l}}\hat{L}%
_{b_{1}}\cdots \hat{L}_{b_{m}}
\end{equation*}%
where $f_{a_{1}\cdots a_{l}}^{b_{1}\cdots b_{m}}$ are constants. This is the
enveloping algebra of the operators of angular momentum and position on a
Hilbert space.

Clearly we see that the $\hat{L}$'s generate the standard angular momentum
algebra and the $\hat{x}$'s transform properly under rotations. However both
the $\hat{x}$'s and the $\hat{L}$'s\ form a constrained version of the
standard $%
\mathbb{R}
^{3}$ Euclidean algebra with invariant constraints given by the last
equations.

\section{Angular Momentum States}

Since we now have the algebra of observables we can ask about Hamiltonians
and states. \ The free single quantum particle Hamiltonian in ordinary
quantum mechanics is $\hat{H}=\frac{\hat{p}^{2}}{2m}=\frac{\hat{p}_{r}^{2}}{%
2m}+\frac{\underline{\hat{L}}\cdot \underline{\hat{L}}}{mr^{2}}$ where $\hat{%
p}_{r}$ is the radial component of momentum and $\underline{\hat{L}}$ is the
angular momentum. \ In other words the natural choice for the Hamiltonian on
our $\mathbb{S}^{2}$ (which we are free to choose) is $\hat{H}=\underline{%
\hat{L}}\cdot \underline{\hat{L}}$, $r=1,m=1\,$\ because it is just the
restricted version of the $\mathbb{E}^{3}$ free particle Hamiltonian onto $%
\mathbb{S}^{2}$. We then construct our angular momentum states in the usual
way by solving the eigenvalue equation:%
\begin{equation}
\hat{H}\left\vert \phi \right\rangle =E\left\vert \phi \right\rangle 
\tag{Schroedinger}  \label{Schroedinger}
\end{equation}%
where $E\in 
\mathbb{R}
$.

We won't do it because it is standard physics that one is able to do as an
undergraduate physics student.

\section{Conclusions}

We have explicitly constructed an exact non-perturbative solutions to the
observables in the Fedosov $\ast $-formalism on $T^{\ast }\mathbb{S}^{2}$
and showed that they obeyed the angular momentum commutation relations. In
other words we took the phase space of a single classical particle confined
to a sphere, quantized it and got the quantum angular momentum algebra
(which we expected). This is done by starting with a chosen phase-space
connection $D$ and constructing an explicit formula for $\hat{D}$. Via the
equation $\left( D-\hat{D}\right) \hat{f}=0$ that defines the algebra i.e.
the algebra of all $\hat{f}$'s we then explicitly constructed $\underline{%
\hat{x}}$ and $\underline{\hat{p}}$ (the operator analogues of $\underline{x}
$ and $\underline{p}$) and computed their commutators. We realized (by
defining $\underline{\hat{L}}=\underline{\hat{x}}$ $\times \underline{\hat{p}%
}$) that the enveloping algebra of all $\underline{\hat{x}}$'s and $%
\underline{\hat{p}}$'s gives the angular momentum algebra.

Subsequently we defined a Hamiltonian $\underline{\hat{L}}\cdot \underline{%
\hat{L}}$ that would have eigenstates of angular momentum, however we did
not explicitly construct it because it is standard physics.

Another main point was that most of the ambiguity given a fixed phase space
connection $D$ of the construction of $\hat{D}$, it seemed, stemmed from the
freedom of a change of basis ($\hat{f}\rightarrow U\hat{f}U^{-1}$) given by
the argument in section 4.1. And finally the matrix form of the $\hat{y}$'s
did not change anything from a Moyal-like object as is done in deformation
quantization.

We conclude that we would arrive at the same answer given any algebraic
object $\hat{y}$ that had the same commutators along with the same action of
the connection on them. We then view the Fedosov $\ast $-formalism as a
general algebraic construction and less tied to the deformation aspect of
its original formulation. Thus our formulation using Heisenberg algebras and
their subsequent representation spaces (Hilbert spaces) makes a more direct
connection to the standard formulation of ordinary quantum mechanics.

\section{Acknowledgements}

We would like to thank E. Ted Newman and Al Janis for their helpful
comments. Also we would like to thank the Laboratory of Axiomatics.

\section{Appendix A}

\hypertarget{A}{}We now show that the equation $\left( D-\hat{D}\right) ^{2}%
\hat{y}^{A}=0$ is equivalent to $\left[ \Omega -Dr+\hat{d}r+r^{2},\hat{y}^{A}%
\right] =0$:

Proof:%
\begin{equation*}
\left( D-\hat{D}\right) ^{2}\hat{y}^{A}=\left( D^{2}-D\hat{D}-\hat{D}D+\hat{D%
}^{2}\right) \hat{y}^{A}
\end{equation*}%
\begin{equation*}
\left( D\hat{D}+\hat{D}D\right) \hat{y}^{A}=\left[ D\left( \omega _{AB}\hat{y%
}^{A}\Theta ^{B}+r\right) ,\hat{y}^{A}\right] =\left[ Dr,\hat{y}^{A}\right]
\end{equation*}%
\begin{eqnarray*}
\hat{D}^{2}\hat{y}^{A} &=&\left[ \hat{Q},\left[ \hat{Q},\hat{y}^{A}\right] %
\right] =\hat{Q}\left( \hat{Q}\hat{y}^{A}-\hat{y}^{A}\hat{Q}\right) +\left( 
\hat{Q}\hat{y}^{A}-\hat{y}^{A}\hat{Q}\right) \hat{Q} \\
&=&\left[ \hat{Q}^{2},\hat{y}^{A}\right] _{-}=\left[ \left( \omega _{AB}\hat{%
y}^{A}\Theta ^{B}+r\right) ^{2},\hat{y}^{A}\right] _{-}=\left[ \left( \omega
_{AB}\hat{y}^{A}\Theta ^{B}\right) ^{2}+\left[ \omega _{AB}\hat{y}^{A}\Theta
^{B},r\right] +r^{2},\hat{y}^{A}\right] _{-}
\end{eqnarray*}%
\begin{equation*}
2\left( \omega _{AB}\hat{y}^{A}\Theta ^{B}\right) ^{2}=\left[ \omega _{AB}%
\hat{y}^{A}\Theta ^{B},\omega _{CE}\hat{y}^{C}\Theta ^{E}\right] =\left[ 
\hat{y}^{A},\hat{y}^{C}\right] \omega _{AB}\Theta ^{B}\omega _{CE}\Theta
^{E}=\omega _{AB}\Theta ^{A}\Theta ^{B}
\end{equation*}%
\begin{equation*}
\implies \hat{D}^{2}\hat{y}^{A}=\left[ \left[ \omega _{AB}\hat{y}^{A}\Theta
^{B},r\right] +r^{2},\hat{y}^{A}\right] _{-}
\end{equation*}%
where $\left[ A,B\right] _{-}=AB-BA$ for any $A$ and $B$.

The curvature $D^{2}$ acting on $\Theta ^{A}$ is:%
\begin{equation*}
D^{2}\otimes \Theta ^{A}=R_{B}^{\text{ \ }A}\otimes \Theta ^{B}
\end{equation*}%
Thus the curvature $D^{2}$ acting on $\hat{y}^{A}$ is: 
\begin{equation*}
D^{2}\hat{y}^{A}=R_{B}^{\text{ \ }A}\hat{y}^{B}
\end{equation*}%
Knowing this we define $\Omega $ as the curvature $D^{2}$ acting on $\hat{y}%
^{A}$ as a commutator, namely:%
\begin{equation*}
\frac{1}{i\hbar }\left[ \Omega ,\hat{y}^{A}\right] =R_{B}^{\text{ \ }A}\hat{y%
}^{B}
\end{equation*}%
we can immediately write a solution for $\Omega $ knowing $\left[ \hat{y}%
^{A},\hat{y}^{B}\right] =i\hbar \omega ^{AB}$, $\omega ^{AB}\omega
_{BC}=\delta _{C}^{A}$ and using the symmetries of the curvature tensor: 
\begin{equation*}
\Omega :=-\frac{1}{2}\omega _{AC}R_{B}^{\text{ \ }A}\hat{y}^{B}\hat{y}^{C}
\end{equation*}%
Thus we may rewrite the condition $\left( D-\hat{D}\right) ^{2}\hat{y}^{A}=0$
as:%
\begin{equation*}
\left( D-\hat{D}\right) ^{2}\hat{y}^{A}=\left[ \Omega -Dr+\hat{d}r+r^{2},%
\hat{y}^{A}\right] =0
\end{equation*}

\section{Appendix B}

\hypertarget{B}{}Here we present an argument as to why $r$ only has terms
that are cubic or higher powers in the $\hat{y}$'s.

Given:%
\begin{equation*}
\hat{D}=\left[ \hat{Q},\cdot \right] =\left[ \hat{Q}_{A}\Theta ^{A},\cdot %
\right]
\end{equation*}%
\begin{equation*}
\hat{Q}_{A}=\sum_{l}Q_{AA_{1}\cdots A_{l}}\hat{y}^{A_{1}}\cdots \hat{y}%
^{A_{l}}
\end{equation*}%
we require:%
\begin{equation*}
\left( D-\hat{D}\right) ^{2}\hat{y}^{A}=0
\end{equation*}%
If we let:%
\begin{equation*}
\hat{Q}_{A}\Theta ^{A}=\omega _{AB}\hat{y}^{A}\Theta ^{B}+r
\end{equation*}%
\begin{equation*}
r=\sum_{l}r_{AA_{1}\cdots A_{l}}\Theta ^{A}\hat{y}^{A_{1}}\cdots \hat{y}%
^{A_{l}}
\end{equation*}%
If we want $r$ to be globally defined for \underline{all} manifolds we must
define it out of non-degenerate tensors namely the metric, the symplectic
form and the curvature. This is because $\Omega $ is degree 2 in the $\hat{y}
$'s (i.e. $\Omega :=-\frac{1}{2}\omega _{AC}R_{B}^{\text{ \ }A}\hat{y}^{B}%
\hat{y}^{C}$ has 2 $\hat{y}$'s). The degree is defined by:%
\begin{equation*}
\deg \left( a\right) =\left( \text{number of }\hat{y}\text{'s}\right)
+2\left( \text{number of }\hbar \text{'s}\right)
\end{equation*}%
A linear $r$ would yield:%
\begin{equation*}
\underset{2}{\underbrace{\Omega }}-\underset{1}{\underbrace{Dr}}+\underset{0}%
{\underbrace{\hat{d}r}}+\underset{1}{\underbrace{r^{2}}}
\end{equation*}%
and this cannot be zero for $\Omega \neq 0$. This means that $r$ must have a
quadratic term in it.

If $r$ is quadratic ($r=\sum_{l=0}^{2}r_{AA_{1}\cdots A_{l}}\Theta ^{A}\hat{y%
}^{A_{1}}\cdots \hat{y}^{A_{l}}$), in general, there is no way to construct
the degree 2 coefficient $r_{AA_{1}A_{2}}$ out of invariant tensors. Thus we
require that $r$ has terms that are cubic or higher powers in the $\hat{y}$%
's. Fedosov mentions this fact also.$\left[ \text{\hyperlink{1}{1}}\right] $

For a specific manifold there might be an $r$ that is quadratic. The
argument above is meant for an $r$ in a \underline{general} construction for
a \underline{general} manifold and so we give a counterexample in the case
when the manifold $\mathcal{M}$ is $\mathbb{E}^{n}$.

There is always the trivial solution to $r$:%
\begin{equation*}
r=-\frac{1}{2}\omega _{CB}\Gamma _{~A}^{C}\hat{y}^{A}\hat{y}^{B}
\end{equation*}%
where $\Gamma _{~A}^{C}=\Gamma _{~BA}^{C}\Theta ^{B}$ are the Christoffel
symbols associated to $D$. One can easily observe that this is a solution
knowing $\left[ \hat{y}^{A},\hat{y}^{B}\right] =i\hbar \omega ^{AB}$, $%
\omega ^{AB}\omega _{BC}=\delta _{C}^{A}$ and using the symmetries of the
Christoffel symbols. However the $\Gamma $'s are not necessarily globally
defined and if we find an $r$ in one coordinate patch on $T^{\ast }\mathcal{M%
}$ there is no guarantee that it will be well-defined in another. However if 
$\mathcal{M}=\mathbb{E}^{n}$ then this is a global $r$.

\section{Appendix C}

\hypertarget{C}{}Useful identities:%
\begin{equation*}
d\underline{p}=\underline{\alpha }\times \underline{x}-\underline{p}\times 
\underline{\theta }
\end{equation*}%
\begin{equation*}
\theta ^{a}\theta ^{b}=\tilde{\omega}\varepsilon ^{abc}x_{c}
\end{equation*}%
\begin{equation*}
\underline{z}\times \underline{x}=\underline{p}\times \underline{x}-%
\underline{k}
\end{equation*}%
\begin{equation*}
\underline{z}=\underline{p}-\underline{x}\times \underline{k}
\end{equation*}%
\begin{equation*}
\theta ^{a}\theta ^{b}=\theta ^{\lbrack a}\theta ^{b]}=\frac{1}{2}%
\varepsilon ^{abc}\left( \underline{\theta }\times \underline{\theta }%
\right) _{c}=\tilde{\omega}\varepsilon ^{abc}x_{c}
\end{equation*}%
\begin{equation*}
\left( \underline{v}\times \underline{w}\right) \times \underline{u}=\delta
_{ab}v^{a}\underline{w}u^{b}-\underline{v}\left( \underline{w}\cdot 
\underline{u}\right)
\end{equation*}%
\begin{equation*}
\underline{v}\times \left( \underline{w}\times \underline{u}\right) =\delta
_{ab}v^{a}\underline{w}u^{b}-\left( \underline{v}\cdot \underline{w}\right) 
\underline{u}
\end{equation*}%
for all 3-D vectors assuming nothing about $\left[ v_{a},w_{b}\right] ,\left[
v_{a},u_{b}\right] $ or $\left[ w_{a},u_{b}\right] $.%
\begin{equation*}
\left( \underline{v}\cdot \underline{\theta }\right) \left( \underline{x}%
\times \underline{w}\right) \cdot \underline{\theta }=\tilde{\omega}\left( 
\underline{v}\cdot \underline{w}\right)
\end{equation*}%
for all 3-D vectors assuming $\left[ \theta ^{a},v_{b}\right] =\left[ \theta
^{a},w_{b}\right] =0$ and assuming nothing about $\left[ v_{a},w_{b}\right] $%
.

For two vectors such that $\underline{v}\cdot \underline{x}=\underline{w}%
\cdot \underline{x}=0$ we have the identities:%
\begin{equation*}
\underline{v}\times \underline{w}=\left( \left( \underline{v}\times 
\underline{w}\right) \cdot \underline{x}\right) \underline{x}\sim \underline{%
x}
\end{equation*}%
\begin{equation*}
\underline{z}\cdot \left( \underline{x}\times \underline{s}\right) =%
\underline{p}\cdot \left( \underline{x}\times \underline{s}\right) -t
\end{equation*}%
\begin{equation*}
\left[ s^{2},\left( \underline{x}\times \underline{k}\right) \cdot 
\underline{s}\right] =0
\end{equation*}%
\begin{equation*}
s_{a}f\left( \underline{k}\cdot \underline{s}\right) =f\left( \underline{k}%
\cdot \underline{s}+1\right) s_{a}
\end{equation*}%
\begin{equation*}
\left[ r_{0},\underline{s}\right] =\frac{1}{3}\left( \left( \underline{s}%
\cdot \underline{\theta }\right) \underline{s}-s^{2}\underline{\theta }%
\right)
\end{equation*}%
\begin{equation*}
\left[ r_{0},\left( \underline{s}\cdot \underline{\theta }\right) \right] =0
\end{equation*}%
\begin{equation*}
\left[ r_{0},s^{2}\right] =0=\left[ \underline{z}\cdot \underline{s},s^{2}%
\right]
\end{equation*}%
\begin{equation*}
\left[ r_{0},\underline{k}\right] =\frac{1}{3}\left( 2\underline{s}\left( 
\underline{k}\cdot \underline{\theta }\right) -\underline{\theta }t-\left( 
\underline{s}\cdot \underline{\theta }\right) \underline{k}\right)
\end{equation*}%
\begin{equation*}
\left[ r_{0},\underline{z}\right] =\frac{1}{3}\left( \left( \underline{s}%
\cdot \underline{\theta }\right) \underline{x}\times \underline{k}-%
\underline{\theta }\times \underline{x}t-2\underline{x}\times \underline{s}%
\left( \underline{k}\cdot \underline{\theta }\right) \right)
\end{equation*}%
\begin{equation*}
\tilde{D}\underline{s}=\underline{\theta }\times \underline{s}-\left( 1+%
\frac{1}{s^{2}}\right) \left( \underline{s}\cdot \underline{\theta }\right) 
\underline{s}-\frac{1}{s^{2}}\left( \left( \underline{x}\times \underline{s}%
\right) \cdot \underline{\theta }\right) \underline{x}\times \underline{s}
\end{equation*}%
\begin{equation*}
\tilde{D}\underline{x}=D\underline{x}=\underline{\theta }\times \underline{x}%
=\frac{1}{s^{2}}\left( \left( \underline{x}\times \underline{s}\right) \cdot 
\underline{\theta }\right) \underline{s}-\left( \underline{s}\cdot 
\underline{\theta }\right) \underline{x}\times \underline{s}
\end{equation*}%
\begin{eqnarray*}
\tilde{D}\underline{z} &=&\underline{\theta }\times \underline{z}+\left(
\left( \underline{z}\cdot \underline{s}\right) \left( \underline{s}\cdot 
\underline{\theta }\right) -\underline{z}\cdot \left( \underline{x}\times 
\underline{s}\right) \left( \left( \underline{x}\times \underline{s}\right)
\cdot \underline{\theta }\right) \right) \frac{1}{s^{2}}\underline{s} \\
&&+2\underline{z}\cdot \left( \underline{x}\times \underline{s}\right)
\left( \underline{s}\cdot \underline{\theta }\right) \frac{1}{s^{2}}%
\underline{x}\times \underline{s}
\end{eqnarray*}

\section{Appendix D: $T^{\ast }%
\mathbb{R}
^{n}$}

\begin{itemize}
\item \hypertarget{D}{}In the case of $T^{\ast }%
\mathbb{R}
^{n}$ we solve equation $\left( \hyperref[r]{r}\right) $ above for $r$ when $%
D\otimes \Theta ^{A}=0$ therefore $D\hat{y}^{A}=0$ and hence $\Omega =0$ and
get the solution $r=0$. This gives us $\hat{D}$ by the formulas $\left( \hat{%
D}\right) $ and $\left( \hat{Q}\right) $: 
\begin{equation*}
\hat{D}=\frac{1}{i\hbar }\left[ \omega _{AB}\hat{y}^{A}\Theta ^{B},\cdot %
\right] =\frac{1}{i\hbar }\left[ \underline{s}\cdot d\underline{p}-%
\underline{k}\cdot d\underline{x},\cdot \right] =\frac{1}{i\hbar }\left[
\left( \underline{x}+\underline{s}\right) \cdot d\underline{p}-\left( 
\underline{p}+\underline{k}\right) \cdot d\underline{x},\cdot \right]
\end{equation*}%
where $s$ and $k$ are the first $n$ $\hat{y}$'s and the last $n$ $\hat{y}$'s
respectively (i.e. $\hat{y}^{A}=\left( s^{a},k_{a}\right) $) also we have $%
\left[ s^{a},s^{b}\right] =0=\left[ k_{a},k_{b}\right] ,~\left[ s^{a},k_{b}%
\right] =i\hbar \delta _{b}^{a}$ and $Ds^{a}=0=Dk_{a}$.

All operators are required to satisfy:%
\begin{equation*}
\frac{\partial \hat{f}}{\partial x^{a}}dx^{a}+\frac{\partial \hat{f}}{%
\partial p_{a}}dp_{a}-\hat{D}\hat{f}=0
\end{equation*}%
\begin{equation*}
\implies \frac{\partial \hat{f}}{\partial x^{a}}dx^{a}+\frac{\partial \hat{f}%
}{\partial p_{a}}dp_{a}=\frac{1}{i\hbar }\left[ \left( \underline{x}+%
\underline{s}\right) \cdot d\underline{p}-\left( \underline{p}+\underline{k}%
\right) \cdot d\underline{x},\hat{f}\right]
\end{equation*}%
This equation is the specific case of the equation $\left( \text{\hyperref[cond fhat%
]{cond $\hat{f}$}}\right) $ for $T^{\ast }%
\mathbb{R}
^{n}$ introduced in section 2.4. The above equation tells us that $\hat{f}$
is a function of $\hat{x}^{a}=x^{a}+s^{a}$ and $\hat{p}_{a}=p_{a}+k_{a}$ ($%
\hat{f}=\hat{f}\left( \hat{x},\hat{p}\right) $) which are solutions to the
equation $\left( \text{\hyperref[cond fhat]{cond $\hat{f}$}}\right) $ i.e.
the coefficients $b_{A_{1}\cdots A_{l}}^{a}\,$\ and $c_{aA_{1}\cdots A_{l}}$
in the case of $T^{\ast }%
\mathbb{R}
^{n}$ introduced in the section 2.4 when $\ell o\left( \hat{f}\right) =x^{a}$
and $\ell o\left( \hat{f}\right) =p_{a}$ respectively. The equation above
implies that $\frac{1}{i\hbar }\left[ \cdot ,\hat{p}_{a}\right] $ generates
the translation on the cotangent bundle in the $x^{a}-$direction and $\frac{1%
}{i\hbar }\left[ \hat{x}^{a},\cdot \right] $ generates the translation on
the cotangent bundle in the $p_{a}-$direction on all observables $\hat{f}$.
See Fedosov for more details on motivating the need for $\hat{D}$.$\left[ 
\text{\hyperlink{1}{1}}\right] \qquad {\tiny \square }$
\end{itemize}

\section{References}

$\left[ \text{\hypertarget{1}{1}}\right] $\ Boris Fedosov,\textit{\
Deformation Quantization and Index Theory,} Akademie, Berlin 1996.

$\left[ \text{\hypertarget{2}{2}}\right] $\ M. Gadella, M. A. del Olmo and
J. Tosiek,\textit{\ Geometrical Origin of the }$\ast -$\textit{product in
the Fedosov Formalism,} \href{http://xxx.lanl.gov/abs/hep-th/0405157}{%
hep-th/0405157v1}.

$\left[ \text{\hypertarget{3}{3}}\right] $\ J. Hancock, M. Walton, B.
Wynder, \textit{Quantum Mechanics Another Way,} \href{http://xxx.lanl.gov/abs/physics/0405029v1%
}{physics/0405029v1}.

$\left[ \text{\hypertarget{4}{4}}\right] $ H. Omori, Y. Maeda and A.
Yoshioka, \textit{Lett. Math. Phys.} \textbf{26}, 285 (1992).

$\left[ \text{\hypertarget{5}{5}}\right] $ A. Connes, M. Flato and D.
Sternheimer, \textit{Lett. Math. Phys.} \textbf{24}, 1 (1992).

$\left[ \text{\hypertarget{6}{6}}\right] $ V. I. Arnold, \textit{%
Mathematical Methods of Classical Mechanics}, Springer, New York 1978.

$\left[ \text{\hypertarget{7}{7}}\right] $ N. Woodhouse, \textit{Geometric
Quantization}, Oxford Univ. Press, New York 1980.

$\left[ \text{\hypertarget{8}{8}}\right] $ J. E. Moyal, Proc. Camb. Phil.
Soc. \textbf{45}, 99 (1949).

$\left[ \text{\hypertarget{9}{9}}\right] $ H. Weyl, \textit{The Theory of
Groups and Quantum Mechanics}, Dover, New York 1931.

$\left[ \text{\hypertarget{10}{10}}\right] $ E. P. Wigner, Phys. Rev. 
\textbf{40}, 749 (1932).

$\left[ \text{\hypertarget{11}{11}}\right] $ H. J. Groenewold, Physica 
\textbf{12}, 405 (1946).

$\left[ \text{\hypertarget{12}{12}}\right] $~P. A. M. Dirac, \textit{The
Principles of Quantum Mechanics}, Oxford Univ. Press, Oxford

1958.

\end{document}